\documentclass[dvips,floatfix,intlimits,pra,showkeys,showpacs,twocolumn]{revtex4}

\usepackage{amsfonts,amsmath,amssymb}
\usepackage{dcolumn}
\usepackage{graphicx}
\usepackage{hyperref}

\begin{document}

\title{Operation of a Stark decelerator with optimum acceptance}%
\author{Ludwig Scharfenberg}%
\email[Author to whom correspondence should be sent. Electronic
mail:~]{ludwig@fhi-berlin.mpg.de}%
\author{Henrik Haak}%
\author{Gerard Meijer}%
\author{Sebastiaan~Y.~T.~van~de~Meerakker}%
\affiliation{Fritz-Haber-Institut der Max-Planck-Gesellschaft,
Faradayweg 4-6,
   14195 Berlin, Germany}%
\date{\today}%
\begin{abstract}\noindent%
With a Stark decelerator, beams of neutral polar molecules can be
accelerated, guided at a constant velocity, or decelerated. The
effectiveness of this process is determined by the 6D volume in
phase space from which molecules are accepted by the Stark
decelerator. Couplings between the longitudinal and transverse
motion of the molecules in the decelerator can reduce this
acceptance. These couplings are nearly absent when the decelerator
operates such that only every third electric field stage is used
for deceleration, while extra transverse focusing is provided by
the intermediate stages. For many applications, the acceptance of
a Stark decelerator in this so-called $s=3$ mode significantly
exceeds that of a decelerator in the conventionally used ($s=1$)
mode. This has been experimentally verified by passing a beam of
OH radicals through a 2.6 meter long Stark decelerator. The
experiments are in quantitative agreement with the results of
trajectory calculations, and can qualitatively be explained with a
simple model for the 6D acceptance. These results imply that the 6D
acceptance of a Stark decelerator in the $s=3$ mode of operation
approaches the optimum value, i.e. the value that is obtained when
any couplings are neglected.
\end{abstract}%
\maketitle%

\section{Introduction}
\label{sec:introduction}

Since its introduction in 1999, the method of Stark deceleration
has developed into an established method for taming molecular
beams \cite{Bethlem:PRL83:1558,Meerakker:NatPhys}. The Stark
deceleration technique combines molecular beam technology with
concepts from charged particle accelerator physics. In essence, a
part of a beam of neutral molecules is selected and decelerated
utilizing the force that polar molecules experience in
inhomogeneous electric fields. A Stark decelerator produces
bunches of state-selected molecules with a computer-controlled
velocity and with narrow velocity distributions. These beams are
ideally suited for a variety of experiments in which the velocity
of the molecules is an important parameter. Applications include
the use of slow molecular beams to enhance the interaction time in
spectroscopic experiments
\cite{Veldhoven:EPJD31:337,Hudson:PRL96:143004} and scattering
studies as a function of the collision energy
\cite{Gilijamse:Science313:1617}. When the molecules are
decelerated to a near standstill, they can be loaded and confined
in traps \cite{Bethlem:Nature406:491,Meerakker:PRL94:023004}. This
allows the observation of molecules in complete isolation for
times up to several seconds, and enables the investigation of
molecular properties in great detail
\cite{Meerakker:PRL95:013003,Hoekstra:PRL98:133001,Gilijamse:JCP127:221102}.

For many of these applications it is crucial that the number
density of the decelerated packets of molecules is further
increased. Higher densities of decelerated molecules will improve
the statistics in metrology experiments and can be decisive, for
instance, for the observation of (in)elastic scattering or
reactive collisions in crossed molecular beam experiments. Higher
densities in the trap are also a prerequisite for the future
application of cooling schemes like evaporative cooling, needed to
reach the regime of degenerate dipolar quantum gases
\cite{Baranov:PhysScriptT102:74}.

The number density of decelerated molecules that can be reached at
the exit of the decelerator is limited by the initial phase space
density in the molecular beam and by the 6D phase space acceptance
of the decelerator. The latter is defined as the volume in 6D
phase space -- the product of the volume in real space and in
velocity space -- from which stable trajectories through the
decelerator originate. In most Stark deceleration experiments to
date, molecular beams with a low initial velocity are slowed down
using decelerators with a rather limited number of electric field
stages. Hence, molecular beams are typically released from a
cooled pulsed valve using Xe or Kr as a carrier gas. The use of Xe
or Kr and the cooling of the pulsed valve strongly enhances
cluster formation, however, and is generally regarded to be
non-ideal for a molecular beam expansion. Moreover, these
decelerators are usually operated at large phase angles. The phase
angle $\phi_0$ determines the deceleration rate per electric field
stage, and ranges from $0^{\circ} < \phi_0 < 90^{\circ}$ for
deceleration, while acceleration occurs from $-90^{\circ} < \phi_0
< 0^{\circ}$. While for increasing values of $\left|\phi_0\right|$ the
deceleration rate gets larger, the longitudinal phase space acceptance gets
smaller.

An obvious route to higher number densities of decelerated
molecules is thus the use of seed gases of lower mass (preferably
Ne or Ar) in room-temperature expansions and the use of low phase
angles in the decelerator. Together this implies, however, that
(much) longer Stark decelerators need to be constructed to
compensate for, both, the higher initial velocity and the lower
deceleration rate. It is not \emph{a priori} clear whether one can
actually transport molecules through such long decelerators
without significant losses. If only molecular trajectories along
the molecular beam axis are considered, the length of the
decelerator is inconsequential, as the deceleration process is
subject to phase stability
\cite{Bethlem:PRL84:5744,Bethlem:PRA65:053416}. In reality the
molecules in the beam have off-axis position and velocity
components, and do not only experience forces in the longitudinal
(forward) direction. The transverse electric field gradients in
the decelerator drive the molecules back towards the molecular
beam axis. The resulting transverse oscillatory motion is strongly
coupled to the longitudinal motion and can result in a reduction
of the 6D phase space acceptance of the decelerator
\cite{Meerakker:PRA73:023401}. Numerical simulations indicate that
this coupled motion does not affect the overall performance of the
relatively short Stark decelerators that have been used and
operated at high phase angles thus far, but that it can severely
affect the performance of longer decelerators that are operated at
low phase angles. In the ideal case, the longitudinal and
transverse motions in the Stark decelerator are completely
uncoupled. This can be achieved by constructing decelerators with
dedicated, spatially separated, elements for focusing and
deceleration \cite{Kalnins:RSI73:2557,Sawyer:EPJD48:197}, as is
common practice in charged particle accelerators
\cite{Lee:AccPhys:1999}. The required electrode geometries make
the decelerator rather complex, however
\cite{Wolfgang:SciAm219:44,Sawyer:EPJD48:197}.

In this paper we exploit that in a Stark decelerator with the
original electrode geometry \cite{Bethlem:PRL83:1558}, the
coupling between the longitudinal and transverse motion can be
significantly reduced when the decelerator is operated in the
so-called $s=3$ mode \cite{Meerakker:PRA71:053409}. In this mode,
only every third electric field stage is used for deceleration,
while extra transverse focusing is provided by the intermediate
stages. We demonstrate and quantify that for many applications,
the acceptance of the Stark decelerator in the $s=3$ mode
significantly exceeds that of a decelerator in the conventionally
used ($s=1$) mode of operation. The improved performance of the
$s=3$ operation mode was demonstrated earlier for guiding at a
constant velocity ($\phi_0 = 0^{\circ}$)
\cite{Meerakker:PRA73:023401} and for deceleration in a relatively
short decelerator at high phase angles \cite{Sawyer:EPJD48:197},
but was not quantified thus far. Here we present experiments in
which a beam of OH radicals passes through a 2.6 meter long Stark
decelerator consisting of 316 electric field stages. In this
machine, the OH radicals can be detected either after 103 or after
316 electric field stages, enabling a direct comparison between
the $s=1$ and $s=3$ modes of operation under otherwise identical
conditions, in particular using the same phase angle $\phi_0$. The
experimental results are in quantitative agreement with the
results of trajectory calculations, and can qualitatively be
explained with a simple model for the 6D acceptance.

\section{Experiment}

\subsection{Experimental set-up}
\begin{figure}[!htb]
    \centering
    \resizebox{\linewidth}{!}
    {\includegraphics[0,0][594,410]{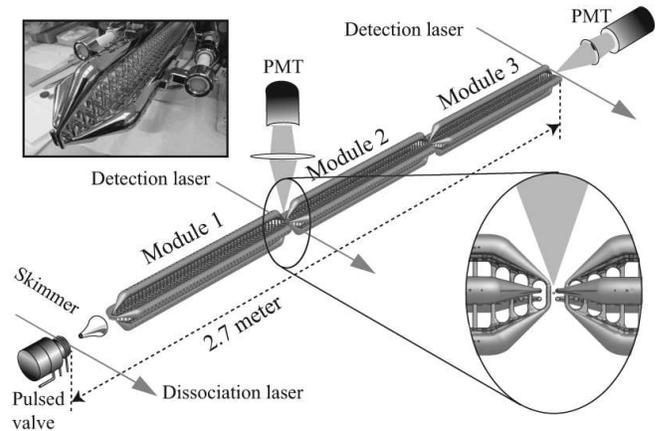}}
    \caption{Scheme of the experimental setup. A pulsed beam of OH radicals
    is produced via photodissociation of HNO$_3$ seeded in Xe, Kr, or Ar.
    The beam of OH radicals passes through a 2.6 meter long Stark decelerator
    that consists of three modules of $\sim$ 100 stages each. The OH radicals
    can be state-selectively detected using a laser induced fluorescence
    scheme at the end of the decelerator, and in the region between the first two
    modules. In the top inset, a photograph of a decelerator module is shown. }
    \label{fig:setup}
\end{figure}

The experimental setup is schematically shown in Figure
\ref{fig:setup}. A pulsed beam of OH radicals is produced via
ArF-laser (193~nm) dissociation of HNO$_3$ seeded in an inert gas.
The dissociation is carried out inside a quartz capillary that is
mounted on the orifice of a pulsed valve (General Valve, Series
99). The experiment runs at a repetition frequency of 10~Hz.
During the supersonic expansion, the majority of the OH radicals
cool to the lowest rotational and vibrational level of the
$X\,^2\Pi_{3/2}$ spin-orbit manifold of the electronic ground
state. This population is equally distributed over the two
$\Lambda$-doublet components of the $J=3/2$ level. Only OH
molecules in the upper $\Lambda$-doublet component are
low-field-seeking, and are of relevance to the experiments
discussed here. This component splits into a $M_J\Omega=-3/4$ and
a $M_J\Omega=-9/4$ component in an electric field. Molecules in
the $M_J\Omega=-9/4$ component experience a Stark shift that is a
factor of three larger than the Stark shift experienced by molecules in
the $M_J\Omega=-3/4$ component.

After passage through a skimmer with a 2~mm diameter opening, the
molecular beam enters the differentially pumped decelerator
chamber. The skimmer is mounted on a compact gate valve
\cite{Kuepper:RSI77:016106}, enabling the venting of the source
chamber while keeping the decelerator chamber under vacuum. The
beam enters the Stark decelerator 60~mm from the nozzle orifice.
The Stark decelerator consists of three modules that are
mechanically and electrically decoupled from each other. The first
two modules consist of 104 electrode pairs (i.e. 103 electric
field stages) each, whereas the last module contains 109 electrode
pairs. These electrode pairs consist of two parallel 4.5~mm
diameter stainless steel electrodes that are placed symmetrically
around the molecular beam axis, providing a 3~mm gap for the
molecular beam to pass through. Adjacent electrode pairs are
alternately horizontally and vertically oriented, such that a $3
\times 3$ mm$^2$ opening area remains for the molecular beam. The
electrodes of all horizontal (vertical) pairs within each module
are electrically connected and switched simultaneously to high
voltage. The center-to-center distance ($L$) of electrodes of
adjacent pairs is 8.25~mm, and the three modules are carefully
aligned to also maintain this distance between the electrode pairs
of adjacent modules. The first and last eight electrode pairs of
each module are mounted on conically shaped rods, as shown
enlarged in Figure \ref{fig:setup}. This design provides excellent
optical access for fluorescence collection in between adjacent
decelerator modules. It also allows the exits of two Stark
decelerators in a crossed beam arrangement to be brought close
together.

\begin{figure}[!htb]
    \centering
    \resizebox{\linewidth}{!}
    {\includegraphics[0,0][490,470]{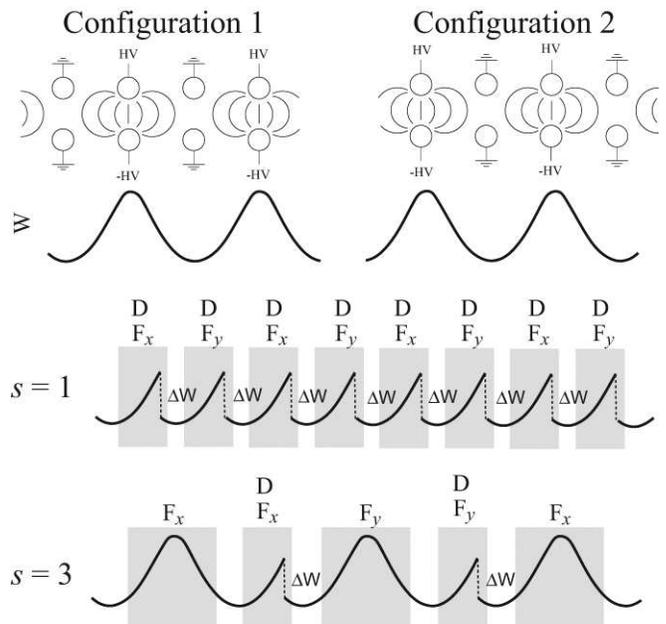}}
    \caption{Schematic representation of the two electric field configurations that are
    used in the deceleration process, together with the potential energy $W$
    for an OH molecule along the molecular beam axis. By switching between the two
    field configurations when the molecules are at the positions indicated by the vertical dashed
    lines, an amount of kinetic energy $\Delta W$ is removed from the molecules.
    In the conventional ($s=1$) mode of operation, each electric field stage is used simultaneously
    for deceleration (D) and focusing in alternating transverse directions ($F_x$,$F_y$). In the
    $s=3$ mode of operation, only every third stage is used for combined deceleration and
    transverse focusing, while the intermediate stages provide additional focusing.}
    \label{fig:scheme}
\end{figure}

The electric field in the decelerator is switched back and forth
between two different configurations, that are schematically shown
in Figure \ref{fig:scheme}. In each configuration, the opposing
electrodes of every other electrode pair are at $\pm 20$~kV, while
the remaining electrodes are grounded. Switching between the two
configurations is performed using fast air-cooled high voltage
switches (Behlke HTS 301-03-GSM). To minimize the power
dissipation per switch, each module is connected to four separate
high voltage switches. Each switch is connected to its power
supply via a 0.5~$\mu$F capacitor bank, limiting the
voltage drop during a deceleration cycle to less than 5~\%.
For the conventional mode of operation of a Stark decelerator, the
$s=1$ mode, the voltages are switched every time the molecules
approach the pair of electrodes that are on high voltage
\cite{Bethlem:PRA65:053416}. In this case, the molecules are
simultaneously decelerated and transversally focussed in every
electric field stage. When the decelerator is operated in the
$s=3$ mode, the voltages are switched only after every third
stage. In this case, only every third stage is used for combined
deceleration and transverse focusing, while the intermediate
stages provide additional focusing.

The OH radicals can be state-selectively detected using an
off-resonant Laser Induced Fluorescence (LIF) detection scheme at
two different positions along the beam line. The first detection
zone is located between the first two modules and the second one
is 18~mm downstream from the last module, enabling the detection
of OH radicals after 103 or 316 electric field stages,
respectively. The 282~nm radiation of a pulsed dye laser crosses
the molecular beam in either one of the detection regions at right
angles, and saturates the (spectroscopically not resolved)
$Q_{21}(1)$ and $Q_1(1)$ transitions of the $A\,^2\Sigma^+, v=1
\leftarrow X\,^2\Pi_{3/2}, v=0$ band. The fluorescence occurs
predominantly on the $A\,^2\Sigma^+,v=1 \rightarrow
X\,^2\Pi, v=1$ transition around 313~nm. Stray light from
the laser is minimized by passing the laser beam through light
baffles between the entrance and exit windows, and by optical
filtering in front of the photomultiplier tube (PMT).

In the experiments the seed gases Xe, Kr, or Ar are used. The mean
initial beam velocities and velocity distributions are deduced
from time-of-flight (TOF) measurements of the OH beam when all
electrode pairs of the decelerator are simultaneously charged to
$\pm$ 7~kV. The mean velocities for the beam are 350~m/s, 430~m/s,
and 590~m/s for Xe, Kr, and Ar, respectively, with a full width at
half maximum (FWHM) velocity spread of about 15-20~\% for all seed
gases.

\begin{figure}[!htb]
    \centering
    \resizebox{\linewidth}{!}
    {\includegraphics[0,0][525,370]{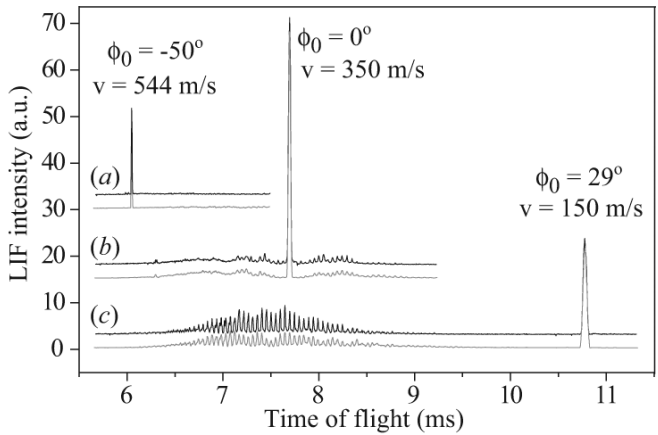}}
    \caption{Time-of-flight profiles of OH radicals, recorded at the exit of the 316 stage Stark decelerator. The Stark decelerator is
    operated at the $s=3$ mode, and accelerates (curve (\emph{a})), guides (curve (\emph{b})), or decelerates (curve (\emph{c})) a
    packet of OH radicals with an initial velocity of 350~m/s. The TOF profiles that result from simulations of the experiment
    are shown underneath the experimental profiles.}
    \label{fig:longTOFs}
\end{figure}

In Figure \ref{fig:longTOFs} the intensity of the LIF signal of a
beam of OH ($J=3/2$) radicals seeded in Xe is shown as a function
of time after firing the dissociation laser using different
deceleration sequences.
The OH radicals are detected using the second LIF detection unit,
and the Stark decelerator is operated in the $s=3$ mode. In curve
(\emph{b}) the TOF profile is shown that is obtained when the
decelerator is operated at a phase angle of $\phi_0=0^{\circ}$,
corresponding to guiding a packet of OH radicals at a constant
velocity. A packet of OH radicals with a mean velocity of 350~m/s
is selected, transported through the 2.6 meter long decelerator,
and arrives in the detection region some 7.6~ms after its
production, with a FWHM of the arrival time distribution of 25
$\mu$s. The measurements shown in curve (\emph{a}) are obtained
with the decelerator operating at a phase angle of
$\phi_0=-50^{\circ}$, accelerating a packet of OH radicals from an
initial velocity of 350~m/s to a final velocity of 544~m/s. There
is no signature of the part of the molecular beam that is not
accelerated. This is also expected as the electrodes of the
decelerator are switched to ground when the accelerated packet
exits the decelerator, about 1.5~ms before the remainder of the
beam pulse would arrive in the detection region. Curve (\emph{c})
shows the TOF profile that is obtained when the decelerator is
operated at a phase angle of $\phi_0=29^{\circ}$ to decelerate a
packet of OH radicals from 350~m/s to 150~m/s. The decelerated
molecules exit the decelerator about 10.7~ms after production,
about 3~ms after the arrival of the undecelerated part of the
beam.

The experimental TOF profiles are in excellent agreement with the
profiles that result from three dimensional trajectory simulations
of the experiment that are shown underneath the experimental
profiles. In these and in all subsequent simulations, the
individual contributions of the $M_J\Omega=-3/4$ and the
$M_J\Omega=-9/4$ components to the LIF signal intensity are taken
into account.

\subsection{Comparing the $s=1$ and $s=3$ modes of operation}\label{sec:s3s1comparison}
\begin{figure*}[!htb]
    \centering
    \resizebox{\linewidth}{!}
    {\includegraphics[0,0][588,284]{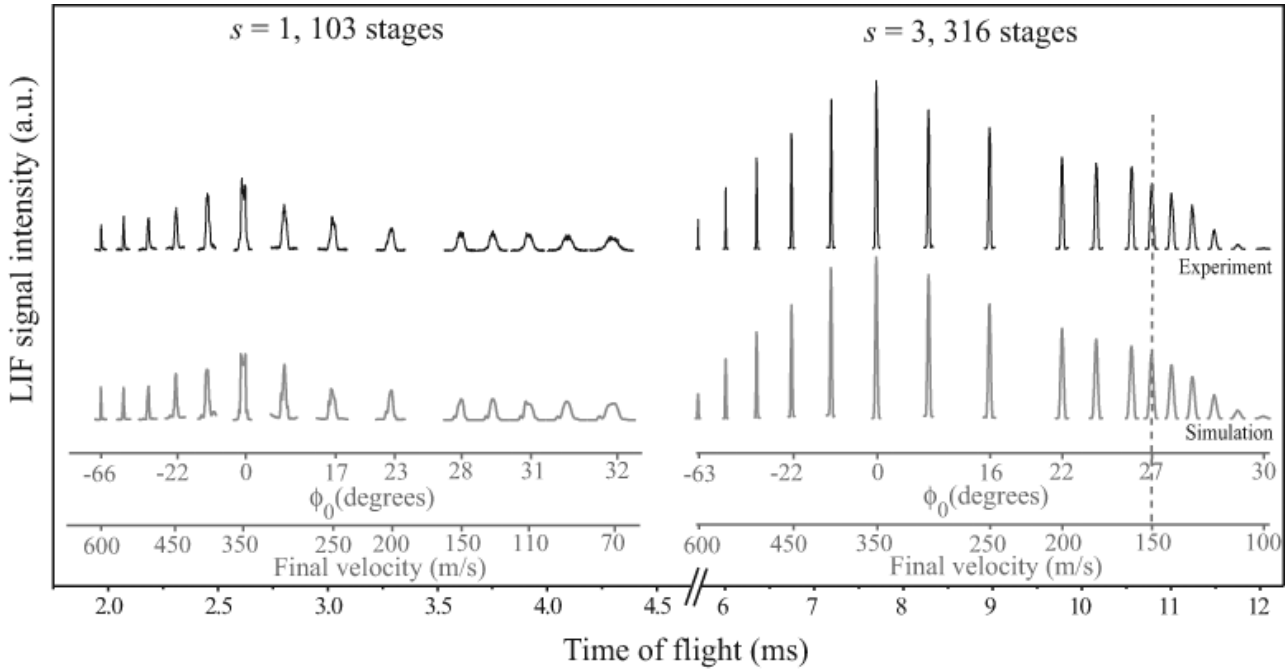}}
    \caption{Time-of-flight profiles of OH radicals that exit the Stark decelerator
    using the $s=1$ and $s=3$ mode of operation. The OH radicals are detected after
    103 and after 316 electric field stages for $s=1$ and $s=3$, respectively. The measurements
    are recorded under otherwise identical conditions, and are shown on the same vertical scale.
    The beam of OH radicals has a mean initial velocity of 350~m/s. The mean final
    velocity of the molecules and the phase angle used, are indicated for selected profiles. }
    \label{fig:comparison-series}
\end{figure*}

In Figure \ref{fig:comparison-series} two series of TOF profiles
are shown that allow a direct comparison between the performance
of a Stark decelerator in the $s=1$ and $s=3$ operation mode under
otherwise identical conditions. In both series, the Stark
decelerator is programmed to accelerate, guide or decelerate a
packet of OH radicals with a mean initial velocity of 350~m/s to a
final velocity that is in the $70-600$~m/s range
($-65^{\circ}<\phi_0<32^{\circ}$). Only that part of each TOF
profile that contains the signature of the packet at the final
velocity is shown. In the left and right panels the series of
profiles are shown that are obtained when the decelerator is
operated using the $s=1$ and the $s=3$ mode, detecting the OH
radicals after 103 and 316 stages, respectively. The (almost)
factor of three difference in the number of stages results in
(almost) identical phase angles for the Stark decelerator to
produce a given final velocity in both series. The phase angle
that is used, and the final velocity of the packet, is indicated
for selected profiles in both panels. To enable a direct
comparison between the two modes of operation, both series are
plotted on the same vertical scale. For this, the relative
detection efficiency in the two LIF zones is experimentally
determined by measuring OH radicals at both detection locations
when the Stark decelerator is operated at $s=3, \phi_0=0^{\circ}$.
The overall scaling factor that is thus determined is correct if
we assume that for the $s=3, \phi_0=0^{\circ}$ mode of operation
the number density of the packet of OH radicals does not decline
when progressing from the first to the second LIF zone, and that
the relative detection efficiency is independent of the velocity
of the OH radicals. The validity of both assumptions is checked
experimentally and verified by numerical simulations.

When the decelerator is operated using $s=1$, the signal intensity
for $\phi_0 = 0^{\circ}$ is about a factor of two-and-a-half lower than
the signal intensity for $s=3, \phi_0=0^{\circ}$. The signal
intensity for the $s=1$ appears rather constant for the different
values of $\phi_0$ that are used. For the $s=3$ mode of operation
it is observed that the signal intensity for $\phi_0 \neq
0^{\circ}$ gradually reduces, following the reduced acceptance of
the decelerator for increasing absolute values of the phase angle.
For final velocities below 150~m/s, indicated by the dashed line
in Figure \ref{fig:comparison-series}, a sharp reduction of the
signal intensity is observed even though the change in phase angle
is only very small. This reduction is due to excessive transverse
focusing for low velocities, and will be discussed in more detail
in section \ref{subsec:excessive}. The signal intensities for the
$s=1$ and $s=3$ modes of operation are about equal for $\phi_0 =
-65^{\circ}$. Both series of TOF profiles are in excellent
agreement with profiles that result from simulations of the
experiment that are shown underneath the experimental profiles. It
is noted that the low field seeking $M_J\Omega=-3/4$ component
only contributes to the TOF profiles for
$-30^{\circ}\leq\phi_0\leq 30^{\circ}$.

\begin{figure}[!htb]
    \centering
    \resizebox{\linewidth}{!}
    {\includegraphics[0,0][456,390]{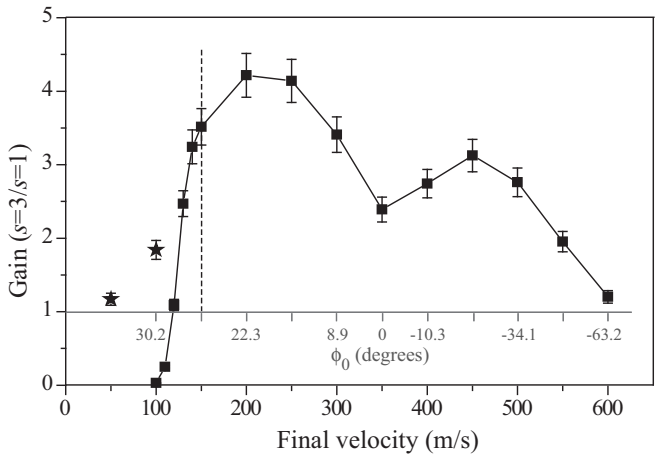}}
    \caption{The ratio of the maximum signal intensities (squares, connected
    by straight line segments) for $s=3$ operation versus $s=1$
    operation as a function of the final velocity.
    The two additional data points (stars) apply to a bimodal operation
    of the decelerator (see text for details).}
    \label{fig:gain}
\end{figure}

The gain of the $s=3$ mode with respect to the $s=1$ mode of
operation, defined as the ratio of the maximum signal intensities
of the decelerated packets at a given final velocity, is shown as
a function of the final velocity in Figure \ref{fig:gain}. For
this, the data presented in Figure \ref{fig:comparison-series} are
used.
For selected velocities, the phase angle that is used for $s=3$
operation is indicated; the phase angle that is used for $s=1$
operation differs only slightly from this value. It is seen that
for $\phi_0=0^{\circ}$, i.e. at a final velocity of 350~m/s, the
gain is 2.4, consistent with previous studies
\cite{Meerakker:PRA71:053409,Sawyer:EPJD48:197}. When the beam is
accelerated to 450~m/s ($\phi_0=-21.7^{\circ}$), a gain up to 3 is
observed. For higher velocities the gain gradually reduces and
reaches 1.2 for 600~m/s ($\phi_0 = -63.2^{\circ}$). A gain up to
4.2 is observed when the beam is decelerated to 200~m/s ($\phi_0 =
22.3^{\circ}$). Below 150~m/s ($\phi_0 = 27.0^{\circ}$), the gain
drops fast and reaches 1.0 for a final velocity of 120~m/s.

\subsection{The $s=1$ mode of operation at low phase angles}\label{pruts}

In this section we want to address the question whether, for a fixed
initial and final velocity, the number density of decelerated molecules
can also be increased by using lower phase angles in the $s=1$ mode
of operation. For this, we compare the deceleration
with 103 stages at a certain phase-angle to the deceleration with
316 stages at about one-third of this phase-angle. The data for
the deceleration at the $s=1$ mode with 103 deceleration stages,
have already been shown and discussed in the previous section. The
complimentary data for the deceleration at the $s=1$ mode with 316
deceleration stages have been measured as well; a beam of OH
radicals with the same initial velocity of 350~m/s is decelerated
or accelerated to the same final velocities between 70~m/s and
600~m/s. The phase-angles used for these measurements are very
low, ranging from $-21.7^{\circ}$ to $10.9^{\circ}$. Again, the
signal intensity that is observed with guiding at the $s=3$ mode,
is used to calibrate the measurements with 316 stages relative to
the ones with 103 stages.

\begin{figure}[!htb]
    \centering
    \resizebox{\linewidth}{!}
    {\includegraphics[0,0][450,350]{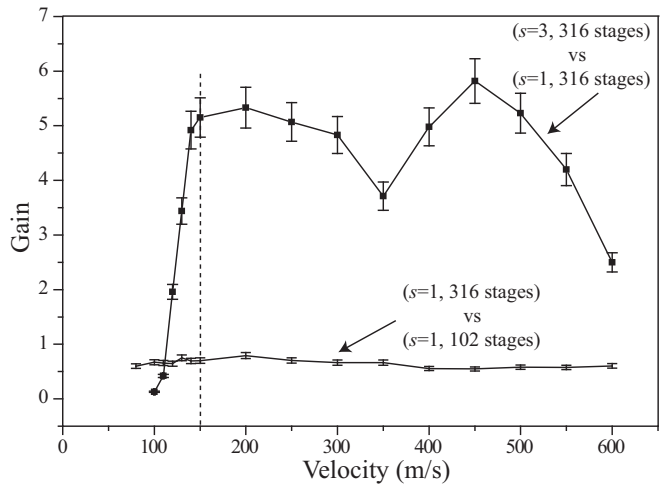}}
    \caption{Lower curve: The ratio of the signal intensity for $s=1$ operation of a decelerator
    with 316 versus 103 stages as a function of the final velocity. Upper curve: The ratio of the
    signal intensity using a 316 stage decelerator operating on the $s=3$ versus the $s=1$ mode.}
    \label{fig:gain-other}
\end{figure}

For the operation in the $s=1$ mode, the gain in using 316 stages
compared to using 103 stages is shown as a function of the final
velocity in the lower curve in Figure \ref{fig:gain-other}. This
gain is determined as the ratio of the signal intensities of the
decelerated packets at a given final velocity, and lies between
0.5 and 0.7 throughout. Since this gain stays smaller than one in
the considered velocity range, it is evident that for the chosen
parameters the number density of decelerated molecules can not be
increased by the use of lower phase-angles. This perhaps
counterintuitive finding is explained by the presence of inherent
instabilities in the $s=1$ mode, which more strongly manifest
themselves during the increased time spent in the decelerator.

We can now also address the question whether, given a Stark
decelerator of a certain length, the maximum number density of
decelerated molecules is obtained using the $s=3$ mode at a
certain phase-angle or using the $s=1$ mode at about one-third of
this phase-angle. For the 316 stage decelerator, the resulting
gain of the $s=3$ versus the $s=1$ mode directly follows from the
curve shown in Figure \ref{fig:gain} and the lower curve in Figure
\ref{fig:gain-other}, and is shown as the upper curve in the
Figure \ref{fig:gain-other}. Operation on the $s=3$ mode is seen
to be about a factor five better than on the $s=1$ mode, provided
the final velocity is larger than the threshold velocity.

\subsection{Excessive focusing at low velocities}\label{subsec:excessive}
The rather abrupt decrease in the number density of decelerated
molecules for velocities below 150~m/s as shown in the right hand
panel of Figure \ref{fig:comparison-series} can qualitatively be
understood as follows. During their flight through the
decelerator, molecules are alternatingly focused in each
transverse direction. When the focusing force acts in one
transverse direction, molecules experience to a good approximation
no focusing or de-focusing force in the orthogonal transverse
direction (the molecules actually experience a small defocusing
force in the orthogonal direction). As long as the characteristic
wavelength $\lambda$ of the transverse oscillatory motion is much
larger than the periodicity of the transverse focusing force,
molecules will follow stable trajectories through the decelerator.
The wavelength $\lambda$ is given by $\lambda=\left<
v_z\right>2\pi/ \Omega_y$, where $\left< v_z\right>$
and $\Omega_y$ are the mean longitudinal velocity
and the mean transverse oscillation frequency of the packet of
molecules, respectively. The periodicity of the focusing force is
given by $2sL$, where $L$ is the center-to-center distance of
adjacent electrode pairs. For high velocities, therefore, stable
trajectories are expected. For low velocities, however, $\lambda$
becomes ever closer to $2sL$, and molecules will get more tightly
transversely focused. The molecular trajectories will then exhibit
ever larger deviations from the molecular beam axis, and the
molecules will eventually crash onto the electrodes. For a given
electric field distribution in a Stark decelerator, the resulting
loss of molecules is thus expected to strongly depend on the
longitudinal velocity of the molecules.

\begin{figure}[!htb]
    \centering
    \resizebox{\linewidth}{!}
    {\includegraphics[0,0][495,700]{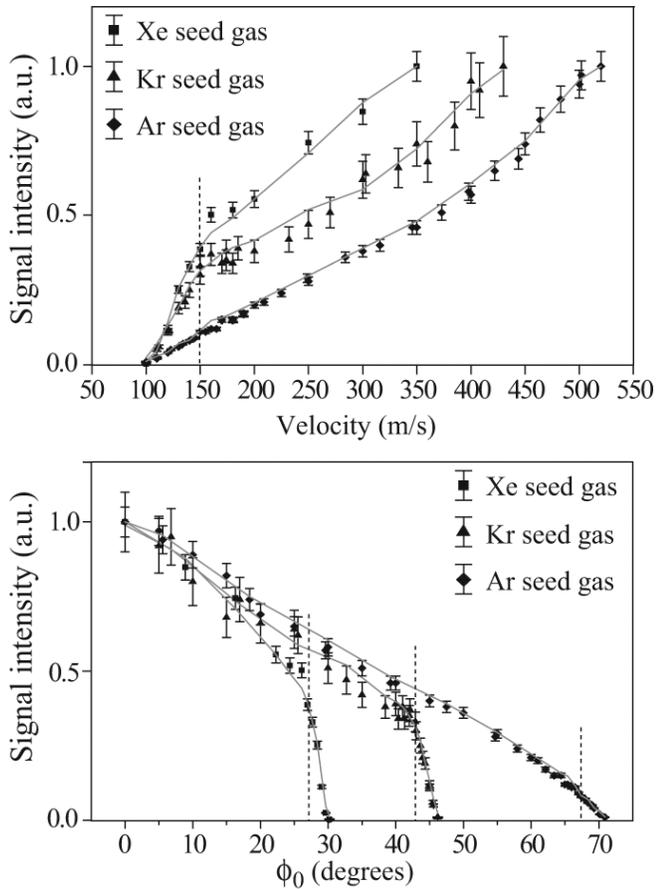}}
    \caption{Maximum signal intensity of decelerated packets of OH radicals as a function
    of the final velocity (upper panel) and as a function of $\phi_0$ (lower panel), using a
    316 stage decelerator operating at the $s=3$ mode. Beams of OH radicals with three
    different initial velocities, produced by seeding in Xe, Kr, or Ar, have been used. The
    intensities that are obtained from numerical simulations of the experiment are shown
    as solid lines.}
    \label{fig:threshold}
\end{figure}

In Figure \ref{fig:threshold}, the maximum signal intensity of
decelerated packets of OH radicals is shown at the exit of the 316
stage decelerator, operating in the $s=3$ mode. Beams of OH
radicals with three different initial velocities have been
used, and the signal is shown as a function of the final velocity
(upper panel) or as a function of the phase-angle (lower panel).
When Xe or Kr are used as carrier gas, the Stark decelerator is
programmed to select a packet of molecules with an initial
velocity that is identical to the mean velocity of the molecular
beam, i.e. 350~m/s or 430~m/s, respectively. For Ar, a velocity of
520~m/s is selected from the slow tail of the velocity
distribution of the beam. The phase angle $\phi_0$ is varied to
decelerate the selected packet of molecules to final velocities
down to 100~m/s. The series of measurements for each seed gas are
normalized to the data point that corresponds to
$\phi_0=0^{\circ}$. The thresholds are found at a velocity of
about 150~m/s in each series, and are indicated by the vertical
dashed lines. This velocity is reached when $\phi_0=27^{\circ}$,
$\phi_0=43^{\circ}$, and $\phi_0=67^{\circ}$ for Xe, Kr, and Ar
seeded beams, respectively. The
value of the threshold velocity appears independent from the phase
angle $\phi_0$ of the decelerator, consistent with the qualitative
picture described above.
The intensities that result from numerical trajectory simulations
of the experiments are shown as solid lines in Figure
\ref{fig:threshold}. Excellent agreement is obtained with the
experiments and in particular the threshold behavior of the signal
intensity is reproduced well.

The threshold velocity below which losses due to excessive
transverse focusing occurs, can be approximately determined as
described in Appendix B. The model described there predicts that
no stable trajectories exist in the $s=3$ mode of operation when $\lambda \leq
(2.6 \cdot 2sL)$. For the present Stark decelerator, operating on
$s=3$, the periodicity of the transverse focusing force $2sL$ is
equal to 49.5~mm. The mean transverse oscillation frequency
$\Omega_y$ follows from the time-averaged transverse
force (see Appendix A). For $s=3$, $\Omega_y$ is
rather independent from the phase angle $\phi_0$, and a threshold velocity
of about 92~m/s is found, consistent with the experimental findings. When
the decelerator is operated in the $s=1$ mode, a threshold
velocity is not so clearly defined, but similar losses due to
excessive transverse focusing occur for velocities below 30~m/s.

The rather high threshold velocity for $s=3$ does not severely
affect experiments in which Stark-decelerated beams are used for
high resolution spectroscopy and collision studies, or in which
the decelerated beams are injected into molecular storage rings or
synchrotrons. It does affect, however, experiments in which lower
final velocities are required, e.g. trap loading experiments.
There are several approaches to yet produce decelerated packets at
a velocity below the threshold velocity with decelerators that are
intended to operate at $s=3$. An electric field geometry for the
last section of the decelerator can be designed that permits a
gradual reduction of the transverse focusing strength. This can be
achieved by a dedicated electrode geometry and/or by a sequential
reduction of the voltage that is applied to the electrodes. It is
noted that similar strategies have already been implemented in
trapping experiments using decelerators in the $s=1$ mode
\cite{Meerakker:ARPC57:159}. An alternative approach is to develop
an electrode geometry for the last segment of the decelerator that
allows the confinement of molecules in a genuine traveling
potential well. When the velocity of this well is gradually
reduced, the packet of molecules can be transferred from the
threshold velocity to lower velocities without loss. The trapping
of molecules in genuine traveling potential wells has already been
demonstrated using optical fields \cite{Fulton:NatPhys2:465} and
using electric fields above a micro-structured electrode array
\cite{Meek:PRL100:153003}.

Within the possibilities of the present experimental arrangement,
decelerated packets with low final velocities can be produced by
changing over from the $s=3$ mode to the $s=1$ mode before the
threshold velocity has been reached. The number of molecules that
exit the decelerator at velocities below the threshold velocity
strongly depends on the details of the change-over, i.e., the velocity
after which $s=1$ operation is used, and the phase angles that are
used before and after the transition. The influence of the choice
of these parameters on the number of molecules that exit the
decelerator has experimentally been studied, decelerating a packet
of OH radicals with an initial velocity of 350~m/s. The velocity
and position of the packet in the decelerator at which the
change-over from $s=3$ to $s=1$ is made is systematically varied
for the target (final) velocities of 100~m/s and 50~m/s. For both
velocities the maximum signal intensity is observed when the
transition to $s=1$ is made when the first $\sim$ 300 stages are
operated at $s=3, \phi_0 \sim 26^{\circ}$ and when the molecular
packet has reached a velocity of 170~m/s. The remaining 12 and 15
stages are then used at $s=1, \phi_0 = 43.2^{\circ}$ and $s=1,
\phi_0 = 48.3^{\circ}$ to produce the final velocities of 100~m/s
and 50~m/s, respectively. The gain of this bimodal operation of
the Stark decelerator with respect to $s=1$ operation is shown in
Figure \ref{fig:gain} as separate data points. It is observed that
for a final velocity of 100~m/s the gain is about a factor of 2,
and is close to one for a final velocity of 50~m/s. These
measurements demonstrate that low final velocities can be produced
with Stark decelerators that are designed to operate at $s=3$,
although the efficiency approaches the efficiency of conventional
Stark decelerators when the target velocity is much lower than the
$s=3$ threshold velocity.

\section{Numerical trajectory calculations}\label{sec:prospects}

The measurements presented thus far demonstrate the performance of
the $s=1$ and $s=3$ modes of operation of Stark decelerators in
the (limited) range of parameters that is accessible to the
experiment. In this section, both modes of operation are studied
in a wider parameter range using numerical trajectory simulations.
The electrode geometry that is used in the simulations is the same
as used in the experiments, but the decelerator is allowed to have
an arbitrary length.

\begin{figure}[!htb]
    \centering
    \resizebox{\linewidth}{!}
    {\includegraphics[0,0][461,570]{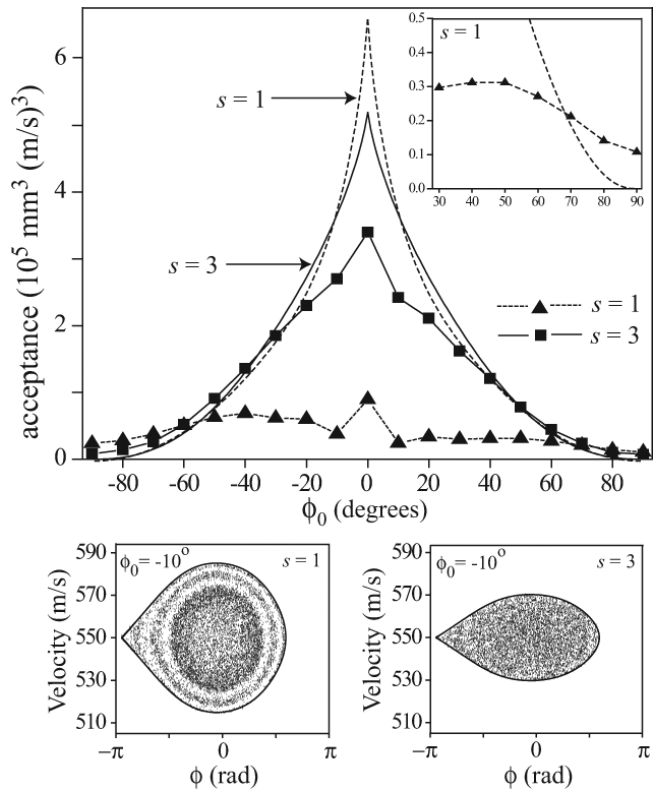}}
    \caption{6D phase space acceptance of a decelerator as a function of the phase-angle $\phi_0$,
    resulting from numerical trajectory calculations (squares connected with solid lines for $s=3$; triangles
    connected with dashed lines for $s=1$), together with the model predictions (solid line for $s=3$; dashed line for $s=1$).
    In the inset, the $s=1$ data is shown enlarged. In the lower part, the longitudinal phase-space distributions that
    result from the simulations are shown for $\phi_0=-10^{\circ}$, both for $s=1$ and $s=3$.}
    \label{fig:comparison}
\end{figure}

Trajectories of OH radicals through the decelerator are
numerically calculated as a function of the phase angle $\phi_0$
for both the $s=1$ and the $s=3$ mode of operation. In these
simulations, a large number of molecules is homogeneously
distributed at the entrance of the Stark decelerator over a block
in 6D phase space. This block has a dimension of 20~mm $\times$
90~m/s in the longitudinal direction, and a dimension of 4~mm
$\times$ 25~m/s in each transverse direction. The molecular
distribution has a mean forward velocity of 550~m/s at the
entrance of the decelerator, and is decelerated or accelerated to
a final velocity of 180~m/s or 755~m/s, corresponding to a change
in the kinetic energy of 90 \%. These values are arbitrary and can
be chosen without loss of generality, as the phase-space
acceptance of a Stark decelerator is in principle independent of
the absolute initial and final velocity. The rather high final
velocity of 180~m/s for deceleration is chosen to stay away from
the velocities for which excessive transverse focusing occurs, as
discussed in section \ref{subsec:excessive}. Decelerators
containing 781, 388, 256, 190, 151, 126, 111, 102, and 99 electric
field stages are simulated that are operated using $|\phi_0| =$
10$^\circ$, 20$^\circ$, 30$^\circ$, 40$^\circ$, 50$^\circ$,
60$^\circ$, 70$^\circ$, 80$^\circ$, and 90$^\circ$, respectively.
These numbers apply to $s=1$; for simulations that apply to $s=3$
the number of stages is three times as large. For both values of
$s$, additional simulations were performed for $\phi_0 =
0^{\circ}$ using a 2500 stages long decelerator. In each
simulation, a sufficient number (5.000.000 for $s=1$ and 500.000
for $s=3$) of molecules is generated to obtain good statistics.
The number of molecules that are within the phase-space
distributions of the decelerated packet are counted, and the
corresponding 6D volume in phase-space is calculated. In the lower
part of Figure \ref{fig:comparison}, the simulated longitudinal
phase-space distributions for $\phi_0=-10^{\circ}$ are shown both
for $s=1$ and for $s=3$, together with the separatrices that
follow from the 1D model for phase stability
\cite{Meerakker:PRA71:053409}. These distributions are
representative for the distributions at low phase angles in
general, and are shown here to exemplify the simulation method
only. The phase-space distribution for $\phi_0=-10^{\circ}, s=1$
is highly structured with alternating stable and unstable regions.
In the distribution for $\phi_0=-10^{\circ}, s=3$ no clear
structure is present. The area within the longitudinal separatrix
-- the longitudinal acceptance -- is a factor $\sqrt{3}$ smaller
for $s=3$ than for $s=1$ \cite{Meerakker:PRA71:053409}.

In the upper panel of Figure \ref{fig:comparison}, the resulting
simulated 6D phase-space acceptance is shown as a function of
$\phi_0$ for both $s=1$ and $s=3$. For $s=1$ the phase-space
acceptance is maximum for $\phi_0 = 0^{\circ}$, and drops
significantly for $\phi_0 \neq 0^{\circ}$. It has a minimum at
$|\phi_0|=10^{\circ}$, is rather constant for $20^{\circ} \leq
|\phi_0| \leq 50^{\circ}$, and drops again for $|\phi_0| \geq
50^{\circ}$, consistent with the experiments in section \ref{pruts}.
Note that the Stark decelerator has a nonzero
acceptance for $|\phi_0| = 90^{\circ}$, as discussed before
\cite{Meerakker:PRA73:023401}. It is interesting to note that the
phase-space acceptance is not symmetric around $\phi_0=0^{\circ}$.
The phase-space acceptance for $\phi_0<0$ is larger than the
acceptance for the same positive phase-angle; molecules spend less
time in the decelerator when they are accelerated then when they
are decelerated, reducing the loss due to instabilities.

The calculated acceptance for $s=3$ shows a rather different
trend, and is generally larger for smaller values of $|\phi_0|$.
Although the longitudinal phase-space acceptance for $s=3$ is a
factor $\sqrt{3}$ smaller than for $s=1$, the 6D phase-space
acceptance exceeds the acceptance for $s=1$ for $0^{\circ} \leq
|\phi_0| \leq 70^{\circ}$. When $|\phi_0| = 70^{\circ}$, the
acceptance in both modes of operation is equal, consistent with
the experimental findings discussed in section
\ref{sec:s3s1comparison}. For $|\phi_0| \geq 70^{\circ}$ the
acceptance for $s=1$ is slightly larger than the acceptance for
$s=3$. Although less pronounced, the $|\phi_0|$ dependence of the
acceptance for $s=3$ is again asymmetric around
$\phi_0=0^{\circ}$.

It is interesting to compare the calculated phase-space acceptance
with the phase-space acceptance that is expected for a Stark
decelerator in which instabilities are absent. From this
comparison one can quantify the presence and severeness of
instabilities in the $s=1$ and $s=3$ mode of operation. The
longitudinal phase-space acceptance of a decelerator is easily
calculated and is, as mentioned above, simply given by the area
within the separatrix
\cite{Bethlem:PRA65:053416,Meerakker:PRA71:053409}. The equations
of motion that govern the transverse trajectories of molecules
through the decelerator contain time-dependent forces, for which
in general no simple analytical solutions exist. These equations,
together with the equation for the longitudinal motion, can be
used to estimate the volume in phase space from which stable
trajectories can originate. This procedure is outlined in detail
in Appendix A, and the resulting 6D phase-space acceptance is
shown as a function of $\phi_0$ for both $s=1$ and $s=3$ in Figure
\ref{fig:comparison}. It is seen that for $s=1$ the phase-space
acceptance predicted by the model deviates significantly from the
calculated acceptance; for $|\phi_0| \leq 20^{\circ}$ the
deviation is at least an order of magnitude. For larger angles the
discrepancy gets less, and both curves cross around
$|\phi_0|=70^{\circ}$. From this comparison it is once more
evident that instabilities are present when the decelerator
operates in the $s=1$ mode \cite{Meerakker:PRA73:023401}, and that
these instabilities severely limit the obtainable acceptance.

For the $s=3$ mode, the acceptance predicted by the model
reproduces the calculated acceptance much better. The agreement
for $|\phi_0|>40^{\circ}$ is good, and the deviations are in the
5-20\% range for $20^{\circ}\leq|\phi_0|\leq 40^{\circ}$ and about 30 \%
for $|\phi_0| < 20^{\circ}$. These minor deviations can be taken
as an indication for the presence of small instable regions, as
have indeed been observed close to the separatrix for $\phi_0 =
0^{\circ}$ \cite{Meerakker:PRA73:023401}. The overall agreement,
however, demonstrates that the 6D acceptance of a Stark
decelerator in the $s=3$ mode of operation approaches the optimum
value, i.e. the value that is predicted from the model that
neglects any instabilities.

\section{Conclusions}

The studies presented in this paper address the question how one
can get the highest number density of decelerated molecules with a
certain velocity at the exit of a decelerator. Rather than discussing a
variety of electrode geometries that one might use to decelerate a beam
of polar molecules, these studies focus on a Stark decelerator in the
conventional, experimentally proven design. This decelerator can run
at different phase-angles and operate in different modes, and can
be built with a variable length. The number density of accelerated
and decelerated OH radicals has been experimentally studied
as a function of these three parameters. Quantitative comparisons of
these number densities, obtained using Stark decelerators with different
parameter sets, have been made. The measurements have been substantiated
by numerical simulations, from which comparisons for a much wider range
of parameters can be made. These studies provide quantitative arguments
for the design criteria of Stark decelerators for specific applications.

Based on the one-dimensional description of a Stark decelerator,
one would expect more molecules at the end of the decelerator for longer
decelerators that run at lower phase-angles. This description neglects the
coupling between the longitudinal and transverse motion, however, which
limits the actual 6D acceptance of a decelerator. A first important conclusion
from the present study is that, for a decelerator operating in the $s=1$
mode, a strategy to optimize the number of decelerated molecules by using
low phase angles and a large number of deceleration stages is only of limited
use. There is a maximum of the 6D acceptance for a phase-angle of around 50$^{\circ}$,
and the optimum number of molecules is obtained when the length of the
decelerator (for a given initial and final velocity) is adjusted such that this
phase-angle can be used. A decelerator of 150 stages that is operated at
50$^{\circ}$, for instance, produces more decelerated molecules than a
decelerator of 250 stages that runs at 30$^{\circ}$.

A second important conclusion is that a decelerator that operates in the
$s=3$ mode outperforms a decelerator in the $s=1$ mode in almost all
cases. In the $s=3$ mode, coupling between the longitudinal and transverse
motion is nearly absent, and lower phase angles always result in a larger
acceptance. For small phase angles, a gain up to a factor of ten can be
obtained. The gain depends strongly on the phase angle that is used and
for phase angles above 70$^{\circ}$, the acceptance for the two modes of
operation is very similar. An intrinsic disadvantage of the $s=3$ mode is
that there are large losses for final velocities below around 150~m/s. Lower
velocities can still be produced, however, and different schemes have been discussed
and demonstrated for this.

A third important conclusion is that the acceptance of a Stark decelerator
operating in the $s=3$ mode approaches the optimum value, i.e. a
decelerator operating in this mode works as good as it can be expected to
get. This conclusion is based on a comparison between the outcome of numerical trajectory
calculations and the 6D acceptance that is derived from a model. In this model,
couplings between the longitudinal and transverse motion are neglected.

To make use of the advantages that the $s=3$ mode of operation
offers, a considerably longer Stark decelerator is needed than for
the $s=1$ mode. This indeed requires more electrode pairs and a
longer vacuum chamber, but it should be realized that there is
no additional requirement on the high voltage electronics.
Compared to the decelerators that have been commonly used so far
($s=1$, $\approx$ 100 stages, $\phi_0 = 50-60^{\circ}$) a five
times longer version operating in the $s=3$ mode at somewhat lower
phase angles will typically result in a factor five higher number
density at the exit. Moreover, this gain in number density is
accompanied by a reduction in the longitudinal translational temperature.

\section{Acknowledgements}
This work is supported by the ESF EuroQUAM programme, and is part of the CoPoMol
(Collisions of Cold Polar Molecules) project. The expert technical assistance of Georg Hammer, Manfred
Erdmann, Rolf Meilicke and the FHI mechanical and electronics workshops, as well
as fruitful discussions with Hendrick L. Bethlem are gratefully acknowledged.

\section*{Appendix A}

In previous work we have studied the transverse stability in a
Stark decelerator \cite{Meerakker:PRA73:023401}. Here we adapt the
model for the transverse motion of molecules through a Stark
decelerator, with the goal to derive the 6D phase-space acceptance
as a function of the phase angle $\phi_0$ that can be expected
when the presence of instabilities is neglected.

In the description of the motion of the OH radicals through the
decelerator, the $z$ coordinate describes the position of the
molecule along the molecular beam axis, while $x$ and $y$ are the
transverse coordinates. The forces in the $x$ and the $y$
direction are assumed to be uncoupled from each other and
identical. The alternating focussing in either one of the
transverse directions (say $y$) is represented by a continuously
acting transverse force $\bar{F_y}(\phi,y)$ that depends on the
phase $\phi$ of the molecule and on $y$ \cite{Meerakker:PRA73:023401}:
\begin{eqnarray}
\nonumber \bar{F}_y (\phi,y)&=& \frac{1}{2T}
\int_t^{t+2T} F_y(y(t'),z(t'),t')
dt' \\
&\approx& \frac{1}{2sL}\int_{\phi L/\pi}^{(\phi+2s\pi)L/\pi}
F_y(y,z) dz,
\end{eqnarray}
where $2T$ is the time during which the synchronous molecule
travels a distance $2sL$.

\begin{figure}[!htb]
    \centering
    \resizebox{0.7\linewidth}{!}
    {\includegraphics[0,-20][480,410]{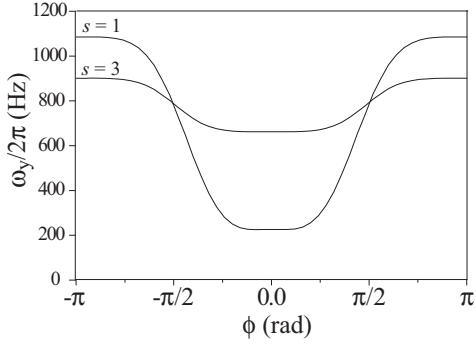}}
    \caption{Natural transverse oscillation frequency $\omega_y/2\pi$
    for an OH ($X\,^2\Pi_{3/2}, J=3/2, M_J\Omega=-9/4$) radical
    as a function of its phase $\phi$ in a Stark decelerator,
    for the operation modes $s=1$ and $s=3$.}
    \label{fig:transverse-frequencies}
\end{figure}

To a good approximation, the transverse force $\bar{F}_y$ is
linear in the displacement $y$ from the molecular beam axis. The
strength of the transverse force can be expressed in terms of a
frequency $\omega_y(\phi)/2\pi$, referred to hereafter as the
natural transverse oscillation frequency, using the relation:
\begin{equation}
\bar{F}_y(\phi,y) = -m \omega^2_y(\phi) \, y,
\end{equation}
where $m$ is the mass of the OH radical. In Figure
\ref{fig:transverse-frequencies}, the natural transverse
oscillation frequency is shown for an OH ($X \,^{2}\Pi_{3/2},
J=3/2, M_J\Omega=-9/4$) radical as a function of its phase $\phi$.
For $s=1$, the natural transverse oscillation frequency has a
strong dependence on the phase $\phi$. For molecules close to
$\phi=\textrm{0}^{\circ}$, the transverse frequency is very low
and focussing forces are almost absent. For $s=3$, the natural
transverse oscillation frequency is rather independent from the
phase $\phi$. The intermediate electric field stage focuses all
molecules equally efficient, effectively decoupling the transverse
focusing properties of the decelerator from the longitudinal phase
$\phi$.

\begin{figure}[!htb]
    \centering
    \resizebox{\linewidth}{!}
    {\includegraphics[0,-50][585,380]{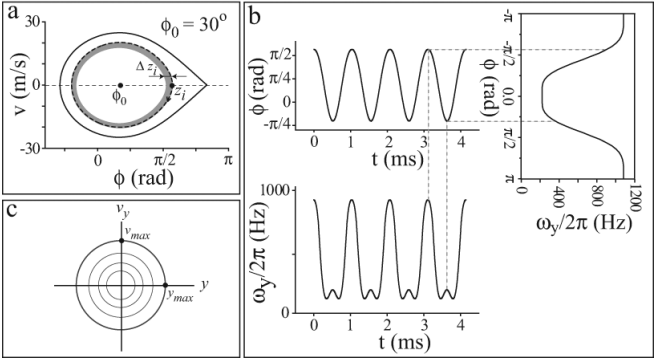}}
    \caption{Schematic representation of the method used to calculate the transverse phase-space acceptance.
    (a) The trajectory of a molecule in longitudinal phase-space, shown as a dashed line. (b) The time dependence
    of the natural transverse oscillation frequency is constructed from the time dependence of the phase $\phi(t)$, and
    the phase dependence of the natural transverse oscillation frequency $\omega_y(\phi)/2\pi$. (c) The time-averaged
    transverse focusing force results in elliptical orbits in transverse phase-space.}
    \label{fig:model-scheme}
\end{figure}

The transverse phase-space acceptance for a given mode of
operation of the decelerator is evaluated as schematically
illustrated in Figure \ref{fig:model-scheme}.
The decelerator is operated at $\phi_0=30^{\circ}$ and $s=1$ in
this example. Let's consider a molecule with a maximum deviation
$z_i$ from the synchronous molecule. The frequency $\omega_y(t)$
of this molecule can be constructed from its phase $\phi(t)$ as it
revolves around the synchronous molecule in longitudinal
phase-space. Let's now consider an ensemble of molecules that is
enclosed by this contour, and by a contour that is displaced by an
infinitesimal value $\Delta z_i$. All these molecules experience
the same temporal dependence of the transverse focusing force, and
the transverse trajectories of the molecules are governed by the
equation:
\begin{equation}
\frac{d^2 y}{dt^2}+\omega_y^2(t) y = 0.
\end{equation}
The transverse phase-space acceptance is easily calculated only if
$\omega^2_y(t)$ is constant. In this case, in which $\omega_y(t)$
is written as $\omega_y$, the longitudinal and transverse motions
are uncoupled, and in each transverse direction the molecules
orbit ellipses in transverse phase-space, as is shown in Figure
\ref{fig:model-scheme}. The phase space acceptance $(A_y)_{z_i}$
and $(A_x)_{z_i}$ in each transverse direction is given by the
maximum extension $y_{max}= x_{max}=$1.5~mm from the molecular
beam axis, and by the maximum transverse velocity $v_{y,max} =
v_{x,max} = \omega_{y}\times y_{max}$ that can be captured. The 4D
volume $(A_t)_{z_i}$ of the transverse phase-space acceptance is
then given by
\begin{equation}\label{eq:trans_phase-space}
(A_t)_{z_i}=(A_x)_{z_i}(A_y)_{z_i} = (\pi \omega_{y}
(y_{max})^2)^2.
\end{equation}
If $\omega^2_y(t)$ is not constant, as is actually the case in a
Stark decelerator, the molecules experience a transverse frequency
$\omega_y(t)$ that oscillates between the minimum value
$\omega_y^{\textrm{min}}$ and the maximum value
$\omega_y^{\textrm{max}}$. The time-averaged value of $\omega^2_y$
for this molecule is given by:
\begin{equation}
\left< \omega^2_y \right>_{z_i} = \frac{1}{\tau}\oint
\omega^2_y(t)dt,
\end{equation}
where $\tau$ is the time it takes the molecule to revolve the contour
in longitudinal phase-space. The transverse phase-space acceptance
cannot be calculated anymore, but the values of
$\omega_y^{\textrm{min}}$, $\omega_y^{\textrm{max}}$, and
$\sqrt{\left< \omega^2_y \right>_{z_i}}$ can nevertheless be used
to characterize the transverse phase-space acceptance of the
ensemble of molecules in three limiting cases. When the
longitudinal oscillation frequency is much larger than the
transverse oscillation frequency, $\omega_y$ can be taken as
$\sqrt{\left< \omega^2_y \right>_{z_i}}$. In that case, one
obtains the 4D transverse acceptance for the molecules in this
shell in longitudinal phase-space. This 4D acceptance can be interpreted as the best estimate for the true
transverse acceptance. When $\omega_y^{\textrm{max}}$ is used, a
transverse acceptances results that can be interpreted as a strict
upper limit for the true acceptance.

The total 6D phase-space acceptance $A(\phi_0)$ is obtained by
integrating over all shells with area $dA_z$ within the separatrix
in longitudinal phase-space:
\begin{equation}
A(\phi_0) = \int(A_y)_{z_i}(A_x)_{z_i} dA_z.
\end{equation}

\begin{figure}[!htb]
    \centering
    \resizebox{\linewidth}{!}
    {\includegraphics[0,-50][390,512]{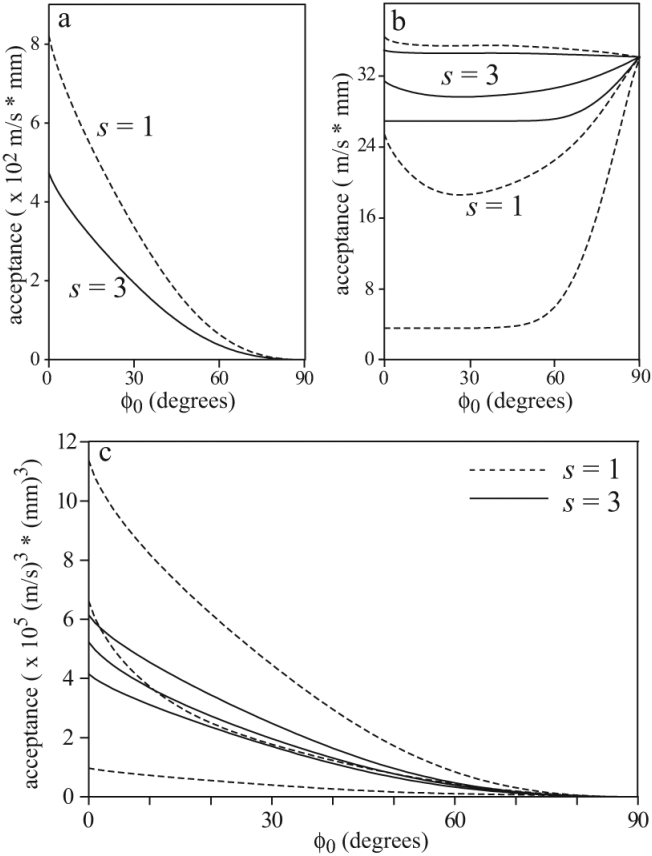}}
    \caption{Prediction for the longitudinal (a), transverse (b) and total 6D (c) phase-space
    acceptance of a Stark decelerator for the operation modes $s=1$ (dashed curves) and $s=3$ (solid curves)
    as a function of the phase angle $\phi_0$. For the transverse and 6D acceptances, three curves are shown
    for each mode of operation that predict the phase space acceptance in three limiting cases, as explained
    in the text. The center curves describe the best estimate for the acceptance in the limit that instabilities
    in the decelerator can be neglected, and have been used in Figure \ref{fig:comparison} of section \ref{sec:prospects}.}
    \label{fig:acceptance}
\end{figure}

In Figure \ref{fig:acceptance}, the longitudinal (2D), transverse
(2D) and total (6D) phase-space acceptances are shown for the
operation modes $s=1$ and $s=3$. The transverse acceptance
$A_y(\phi_0)$ is calculated from the total 6D acceptance
$A(\phi_0)$ and the longitudinal acceptance $A_z(\phi_0)$ via
$A_y(\phi_0)=\sqrt{A(\phi_0)/A_z(\phi_0)}$. For each mode of
operation three curves are shown; the lower, center, and upper
curve correspond to the choice of $\omega_y^{\textrm{min}}$,
$\sqrt{\left< \omega^2_y \right>_{z_i}}$, and
$\omega_y^{\textrm{max}}$ for $\omega_y$ in equation
(\ref{eq:trans_phase-space}), respectively.
The longitudinal acceptance $A_z(\phi_0)$ is a factor $\sqrt{3}$
smaller for $s=3$ than for $s=1$. The transverse acceptance for
$s=3$ is somewhat larger than for $s=1$ and is almost independent
of $\phi_0$. Furthermore, the three different curves predict a
rather similar transverse acceptance for $s=3$. The three curves
for the transverse acceptance for $s=1$ differ much more among
each other. The center curves that are shown in Figure
\ref{fig:acceptance}(c), and that have been used in Figure
\ref{fig:comparison} of section \ref{sec:prospects}, predict a
rather similar total 6D phase-space acceptance for $s=1$ and
$s=3$. For almost all phase angles, the lower longitudinal
phase-space acceptance for $s=3$ is compensated for by the larger
transverse acceptance of $s=3$.

\section*{Appendix B}

To obtain a more quantitative value for the threshold velocity
below which excessive focusing occurs in the $s=3$ mode,
we model the motion of a molecule with a longitudinal velocity that
corresponds to the mean longitudinal velocity $\left< v_z \right>$ of the
ensemble. This molecule experiences a time-averaged harmonic transverse
force with a force constant $m\left< \omega^2_y \right>$. This
force constant is deduced from the transverse acceptance $A_y(\phi_0)$
(center curve for $s=3$ in Figure \ref{fig:acceptance}(b) in Appendix A)
via $m\left<\omega^2_y \right> = m A^2_y(\phi_0)/(\pi y_{max}^2)^2$, and can be
interpreted as the weighed average force constant of the ensemble of molecules
in the packet.

The transverse force on a molecule in the decelerator has periodicity $2sL$,
and is modeled here by a square wave. In a given transverse direction, the
molecule repeatedly experiences a constant transverse focusing force with
harmonic frequency $\Omega_y$ followed by a section of free flight.
The sections for focusing and free flight have equal length $sL$, such that
$\Omega^2_y$ = 2$\left< \omega^2_y \right>$.

For a constant velocity $\left< v_z \right>$, the resulting equation of motion is of the type
of Hill's differential equation, and the stability of transverse trajectories can be
predicted with the transfer matrix method. The transformation matrices of the
coordinate $y$ and velocity $v_y$ for the free flight and focusing sections are given by
\begin{equation}
{y_f \choose v_{y_f}} = \left( \begin{array}{cc} 1 & T\\
0 & 1
\end{array} \right) {y_i \choose v_{y_i}}
\end{equation}
and
\begin{equation}
{y_f \choose v_{y_f}} = \left( \begin{array}{cc} \cos(\Omega_y T)  & \frac{1}{\Omega_y }\sin(\Omega_y T) \\
-\Omega_y \sin(\Omega_y T) & \cos(\Omega_y T)
\end{array} \right) {y_i \choose v_{y_i}},
\end{equation}
respectively, where $T=sL/\left< v_z \right>$ and where the
subscripts $i$ and $f$ refer to initial and final coordinates,
respectively. The transformation matrix $M$ for one full period is
obtained by multiplication of the free flight and focusing
matrices. The trajectories of the molecules through an infinite
number of these periods are stable if $-2 \leq Tr(M) \leq 2$,
where $Tr(M)$ is the trace of the transformation matrix
\cite{Courant:AnnPhys3:1}. When $Tr(M)$ is outside this range, the
amplitude of the transverse motion grows without bound.

In a Stark decelerator, the mean velocity $\left< v_z \right>$ of the packet of molecules is not constant but is gradually changed.
As long as the velocity change per electric field stage is modest compared to the longitudinal velocity, the assumptions
in the model hold. For phase-angles in the range of 30-40$^{\circ}$, $\Omega_y$ is found as $\sqrt{2} \cdot 2\pi \cdot 720$~Hz.
For this value of $\Omega_y$, the inequality
\begin{equation}
\left|2\cos(\Omega_y T) - \Omega_y T\sin(\Omega_y T) \right|\leq 2
\end{equation}
implies that transverse trajectories are not stable for velocities $\left< v_z \right>$
below 92~m/s. Or, in other words, the characteristic wavelength $\lambda$ for the transverse
oscillation, given by $\lambda = 2\pi \left< v_z \right>/\Omega_y$, has to be larger than $2.6 \cdot 2sL$.


\begin{thebibliography}{24}
\expandafter\ifx\csname
natexlab\endcsname\relax\def\natexlab#1{#1}\fi
\expandafter\ifx\csname bibnamefont\endcsname\relax
  \def\bibnamefont#1{#1}\fi
\expandafter\ifx\csname bibfnamefont\endcsname\relax
  \def\bibfnamefont#1{#1}\fi
\expandafter\ifx\csname citenamefont\endcsname\relax
  \def\citenamefont#1{#1}\fi
\expandafter\ifx\csname url\endcsname\relax
  \def\url#1{\texttt{#1}}\fi
\expandafter\ifx\csname
urlprefix\endcsname\relax\def\urlprefix{URL }\fi
\providecommand{\bibinfo}[2]{#2}
\providecommand{\eprint}[2][]{\url{#2}}

\bibitem[{\citenamefont{Bethlem et~al.}(1999)\citenamefont{Bethlem, Berden, and
  Meijer}}]{Bethlem:PRL83:1558}
\bibinfo{author}{\bibfnamefont{H.~L.} \bibnamefont{Bethlem}},
  \bibinfo{author}{\bibfnamefont{G.}~\bibnamefont{Berden}}, \bibnamefont{and}
  \bibinfo{author}{\bibfnamefont{G.}~\bibnamefont{Meijer}},
  \bibinfo{journal}{Phys. Rev. Lett.} \textbf{\bibinfo{volume}{83}},
  \bibinfo{pages}{1558} (\bibinfo{year}{1999}).

\bibitem[{\citenamefont{van~de Meerakker et~al.}(2008)\citenamefont{van~de
  Meerakker, Bethlem, and Meijer}}]{Meerakker:NatPhys}
\bibinfo{author}{\bibfnamefont{S.~Y.~T.} \bibnamefont{van~de Meerakker}},
  \bibinfo{author}{\bibfnamefont{H.~L.} \bibnamefont{Bethlem}},
  \bibnamefont{and} \bibinfo{author}{\bibfnamefont{G.}~\bibnamefont{Meijer}},
  \bibinfo{journal}{Nature Physics} \textbf{\bibinfo{volume}{in press}}
  (\bibinfo{year}{2008}).

\bibitem[{\citenamefont{van Veldhoven et~al.}(2004)\citenamefont{van Veldhoven,
  K\"upper, Bethlem, Sartakov, van Roij, and Meijer}}]{Veldhoven:EPJD31:337}
\bibinfo{author}{\bibfnamefont{J.}~\bibnamefont{van Veldhoven}},
  \bibinfo{author}{\bibfnamefont{J.}~\bibnamefont{K\"upper}},
  \bibinfo{author}{\bibfnamefont{H.~L.} \bibnamefont{Bethlem}},
  \bibinfo{author}{\bibfnamefont{B.}~\bibnamefont{Sartakov}},
  \bibinfo{author}{\bibfnamefont{A.~J.~A.} \bibnamefont{van Roij}},
  \bibnamefont{and} \bibinfo{author}{\bibfnamefont{G.}~\bibnamefont{Meijer}},
  \bibinfo{journal}{Eur. Phys. J. D}
  \textbf{\bibinfo{volume}{31}}(\bibinfo{number}{2}), \bibinfo{pages}{337}
  (\bibinfo{year}{2004}).

\bibitem[{\citenamefont{Hudson et~al.}(2006)\citenamefont{Hudson, Lewandowski,
  Sawyer, and Ye}}]{Hudson:PRL96:143004}
\bibinfo{author}{\bibfnamefont{E.~R.} \bibnamefont{Hudson}},
  \bibinfo{author}{\bibfnamefont{H.~J.} \bibnamefont{Lewandowski}},
  \bibinfo{author}{\bibfnamefont{B.~C.} \bibnamefont{Sawyer}},
  \bibnamefont{and} \bibinfo{author}{\bibfnamefont{J.}~\bibnamefont{Ye}},
  \bibinfo{journal}{Phys. Rev. Lett.}
  \textbf{\bibinfo{volume}{96}}(\bibinfo{number}{14}), \bibinfo{pages}{143004}
  (\bibinfo{year}{2006}).

\bibitem[{\citenamefont{Gilijamse et~al.}(2006)\citenamefont{Gilijamse,
  Hoekstra, van~de Meerakker, Groenenboom, and
  Meijer}}]{Gilijamse:Science313:1617}
\bibinfo{author}{\bibfnamefont{J.~J.} \bibnamefont{Gilijamse}},
  \bibinfo{author}{\bibfnamefont{S.}~\bibnamefont{Hoekstra}},
  \bibinfo{author}{\bibfnamefont{S.~Y.~T.} \bibnamefont{van~de Meerakker}},
  \bibinfo{author}{\bibfnamefont{G.~C.} \bibnamefont{Groenenboom}},
  \bibnamefont{and} \bibinfo{author}{\bibfnamefont{G.}~\bibnamefont{Meijer}},
  \bibinfo{journal}{Science} \textbf{\bibinfo{volume}{313}},
  \bibinfo{pages}{1617} (\bibinfo{year}{2006}).

\bibitem[{\citenamefont{Bethlem
  et~al.}(2000{\natexlab{a}})\citenamefont{Bethlem, Berden, Crompvoets, Jongma,
  van Roij, and Meijer}}]{Bethlem:Nature406:491}
\bibinfo{author}{\bibfnamefont{H.~L.} \bibnamefont{Bethlem}},
  \bibinfo{author}{\bibfnamefont{G.}~\bibnamefont{Berden}},
  \bibinfo{author}{\bibfnamefont{F.~M.~H.} \bibnamefont{Crompvoets}},
  \bibinfo{author}{\bibfnamefont{R.~T.} \bibnamefont{Jongma}},
  \bibinfo{author}{\bibfnamefont{A.~J.~A.} \bibnamefont{van Roij}},
  \bibnamefont{and} \bibinfo{author}{\bibfnamefont{G.}~\bibnamefont{Meijer}},
  \bibinfo{journal}{Nature} \textbf{\bibinfo{volume}{406}},
  \bibinfo{pages}{491} (\bibinfo{year}{2000}{\natexlab{a}}).

\bibitem[{\citenamefont{van~de Meerakker
  et~al.}(2005{\natexlab{a}})\citenamefont{van~de Meerakker, Smeets, Vanhaecke,
  Jongma, and Meijer}}]{Meerakker:PRL94:023004}
\bibinfo{author}{\bibfnamefont{S.~Y.~T.} \bibnamefont{van~de Meerakker}},
  \bibinfo{author}{\bibfnamefont{P.~H.~M.} \bibnamefont{Smeets}},
  \bibinfo{author}{\bibfnamefont{N.}~\bibnamefont{Vanhaecke}},
  \bibinfo{author}{\bibfnamefont{R.~T.} \bibnamefont{Jongma}},
  \bibnamefont{and} \bibinfo{author}{\bibfnamefont{G.}~\bibnamefont{Meijer}},
  \bibinfo{journal}{Phys. Rev. Lett.} \textbf{\bibinfo{volume}{94}},
  \bibinfo{pages}{023004} (\bibinfo{year}{2005}{\natexlab{a}}).

\bibitem[{\citenamefont{van~de Meerakker
  et~al.}(2005{\natexlab{b}})\citenamefont{van~de Meerakker, Vanhaecke, van~der
  Loo, Groenenboom, and Meijer}}]{Meerakker:PRL95:013003}
\bibinfo{author}{\bibfnamefont{S.~Y.~T.} \bibnamefont{van~de Meerakker}},
  \bibinfo{author}{\bibfnamefont{N.}~\bibnamefont{Vanhaecke}},
  \bibinfo{author}{\bibfnamefont{M.~P.~J.} \bibnamefont{van~der Loo}},
  \bibinfo{author}{\bibfnamefont{G.~C.} \bibnamefont{Groenenboom}},
  \bibnamefont{and} \bibinfo{author}{\bibfnamefont{G.}~\bibnamefont{Meijer}},
  \bibinfo{journal}{Phys. Rev. Lett.}
  \textbf{\bibinfo{volume}{95}}(\bibinfo{number}{1}), \bibinfo{pages}{013003}
  (\bibinfo{year}{2005}{\natexlab{b}}).

\bibitem[{\citenamefont{Hoekstra et~al.}(2007)\citenamefont{Hoekstra,
  Gilijamse, Sartakov, Vanhaecke, Scharfenberg, van~de Meerakker, and
  Meijer}}]{Hoekstra:PRL98:133001}
\bibinfo{author}{\bibfnamefont{S.}~\bibnamefont{Hoekstra}},
  \bibinfo{author}{\bibfnamefont{J.~J.} \bibnamefont{Gilijamse}},
  \bibinfo{author}{\bibfnamefont{B.}~\bibnamefont{Sartakov}},
  \bibinfo{author}{\bibfnamefont{N.}~\bibnamefont{Vanhaecke}},
  \bibinfo{author}{\bibfnamefont{L.}~\bibnamefont{Scharfenberg}},
  \bibinfo{author}{\bibfnamefont{S.~Y.~T.} \bibnamefont{van~de Meerakker}},
  \bibnamefont{and} \bibinfo{author}{\bibfnamefont{G.}~\bibnamefont{Meijer}},
  \bibinfo{journal}{Phys. Rev. Lett.}
  \textbf{\bibinfo{volume}{98}}(\bibinfo{number}{13}), \bibinfo{eid}{133001}
  (\bibinfo{year}{2007}).

\bibitem[{\citenamefont{Gilijamse et~al.}(2007)\citenamefont{Gilijamse,
  Hoekstra, Meek, Mets\"al\"a, van~de Meerakker, Meijer, and
  Groenenboom}}]{Gilijamse:JCP127:221102}
\bibinfo{author}{\bibfnamefont{J.~J.} \bibnamefont{Gilijamse}},
  \bibinfo{author}{\bibfnamefont{S.}~\bibnamefont{Hoekstra}},
  \bibinfo{author}{\bibfnamefont{S.~A.} \bibnamefont{Meek}},
  \bibinfo{author}{\bibfnamefont{M.}~\bibnamefont{Mets\"al\"a}},
  \bibinfo{author}{\bibfnamefont{S.~Y.~T.} \bibnamefont{van~de Meerakker}},
  \bibinfo{author}{\bibfnamefont{G.}~\bibnamefont{Meijer}}, \bibnamefont{and}
  \bibinfo{author}{\bibfnamefont{G.~C.} \bibnamefont{Groenenboom}},
  \bibinfo{journal}{J. Chem. Phys.} \textbf{\bibinfo{volume}{127}},
  \bibinfo{pages}{221102} (\bibinfo{year}{2007}).

\bibitem[{\citenamefont{Baranov et~al.}(2002)\citenamefont{Baranov, Dobrek,
  G\'oral, Santos, and Lewenstein}}]{Baranov:PhysScriptT102:74}
\bibinfo{author}{\bibfnamefont{M.}~\bibnamefont{Baranov}},
  \bibinfo{author}{\bibfnamefont{L.}~\bibnamefont{Dobrek}},
  \bibinfo{author}{\bibfnamefont{K.}~\bibnamefont{G\'oral}},
  \bibinfo{author}{\bibfnamefont{L.}~\bibnamefont{Santos}}, \bibnamefont{and}
  \bibinfo{author}{\bibfnamefont{M.}~\bibnamefont{Lewenstein}},
  \bibinfo{journal}{Phys. Scr.} \textbf{\bibinfo{volume}{T102}},
  \bibinfo{pages}{74} (\bibinfo{year}{2002}).

\bibitem[{\citenamefont{Bethlem
  et~al.}(2000{\natexlab{b}})\citenamefont{Bethlem, Berden, van Roij,
  Crompvoets, and Meijer}}]{Bethlem:PRL84:5744}
\bibinfo{author}{\bibfnamefont{H.~L.} \bibnamefont{Bethlem}},
  \bibinfo{author}{\bibfnamefont{G.}~\bibnamefont{Berden}},
  \bibinfo{author}{\bibfnamefont{A.~J.~A.} \bibnamefont{van Roij}},
  \bibinfo{author}{\bibfnamefont{F.~M.~H.} \bibnamefont{Crompvoets}},
  \bibnamefont{and} \bibinfo{author}{\bibfnamefont{G.}~\bibnamefont{Meijer}},
  \bibinfo{journal}{Phys. Rev. Lett.} \textbf{\bibinfo{volume}{84}},
  \bibinfo{pages}{5744} (\bibinfo{year}{2000}{\natexlab{b}}).

\bibitem[{\citenamefont{Bethlem et~al.}(2002)\citenamefont{Bethlem, Crompvoets,
  Jongma, van~de Meerakker, and Meijer}}]{Bethlem:PRA65:053416}
\bibinfo{author}{\bibfnamefont{H.~L.} \bibnamefont{Bethlem}},
  \bibinfo{author}{\bibfnamefont{F.~M.~H.} \bibnamefont{Crompvoets}},
  \bibinfo{author}{\bibfnamefont{R.~T.} \bibnamefont{Jongma}},
  \bibinfo{author}{\bibfnamefont{S.~Y.~T.} \bibnamefont{van~de Meerakker}},
  \bibnamefont{and} \bibinfo{author}{\bibfnamefont{G.}~\bibnamefont{Meijer}},
  \bibinfo{journal}{Phys. Rev. A}
  \textbf{\bibinfo{volume}{65}}(\bibinfo{number}{5}), \bibinfo{pages}{053416}
  (\bibinfo{year}{2002}).

\bibitem[{\citenamefont{van~de Meerakker
  et~al.}(2006{\natexlab{a}})\citenamefont{van~de Meerakker, Vanhaecke,
  Bethlem, and Meijer}}]{Meerakker:PRA73:023401}
\bibinfo{author}{\bibfnamefont{S.~Y.~T.} \bibnamefont{van~de Meerakker}},
  \bibinfo{author}{\bibfnamefont{N.}~\bibnamefont{Vanhaecke}},
  \bibinfo{author}{\bibfnamefont{H.~L.} \bibnamefont{Bethlem}},
  \bibnamefont{and} \bibinfo{author}{\bibfnamefont{G.}~\bibnamefont{Meijer}},
  \bibinfo{journal}{Phys. Rev. A}
  \textbf{\bibinfo{volume}{73}}(\bibinfo{number}{2}), \bibinfo{pages}{023401}
  (\bibinfo{year}{2006}{\natexlab{a}}).

\bibitem[{\citenamefont{Kalnins et~al.}(2002)\citenamefont{Kalnins, Lambertson,
  and Gould}}]{Kalnins:RSI73:2557}
\bibinfo{author}{\bibfnamefont{J.~G.} \bibnamefont{Kalnins}},
  \bibinfo{author}{\bibfnamefont{G.}~\bibnamefont{Lambertson}},
  \bibnamefont{and} \bibinfo{author}{\bibfnamefont{H.}~\bibnamefont{Gould}},
  \bibinfo{journal}{Rev. Sci. Instrum.} \textbf{\bibinfo{volume}{73}},
  \bibinfo{pages}{2557} (\bibinfo{year}{2002}).

\bibitem[{\citenamefont{Sawyer et~al.}(2008)\citenamefont{Sawyer, Stuhl, Lev,
  Ye, and Hudson}}]{Sawyer:EPJD48:197}
\bibinfo{author}{\bibfnamefont{B.~C.} \bibnamefont{Sawyer}},
  \bibinfo{author}{\bibfnamefont{B.~K.} \bibnamefont{Stuhl}},
  \bibinfo{author}{\bibfnamefont{B.~L.} \bibnamefont{Lev}},
  \bibinfo{author}{\bibfnamefont{J.}~\bibnamefont{Ye}}, \bibnamefont{and}
  \bibinfo{author}{\bibfnamefont{E.~R.} \bibnamefont{Hudson}},
  \bibinfo{journal}{Eur. Phys. J. D} \textbf{\bibinfo{volume}{48}},
  \bibinfo{pages}{197} (\bibinfo{year}{2008}).

\bibitem[{\citenamefont{Lee}(1999)}]{Lee:AccPhys:1999}
\bibinfo{author}{\bibfnamefont{S.~Y.} \bibnamefont{Lee}},
  \emph{\bibinfo{title}{Accelerator physics}} (\bibinfo{publisher}{World
  Scientific}, \bibinfo{address}{Singapore}, \bibinfo{year}{1999}), ISBN
  \bibinfo{isbn}{981023709X}.

\bibitem[{\citenamefont{Wolfgang}(1968)}]{Wolfgang:SciAm219:44}
\bibinfo{author}{\bibfnamefont{R.}~\bibnamefont{Wolfgang}},
  \bibinfo{journal}{Sci. Am.}
  \textbf{\bibinfo{volume}{219}}(\bibinfo{number}{4}), \bibinfo{pages}{44}
  (\bibinfo{year}{1968}).

\bibitem[{\citenamefont{van~de Meerakker
  et~al.}(2005{\natexlab{c}})\citenamefont{van~de Meerakker, Vanhaecke,
  Bethlem, and Meijer}}]{Meerakker:PRA71:053409}
\bibinfo{author}{\bibfnamefont{S.~Y.~T.} \bibnamefont{van~de Meerakker}},
  \bibinfo{author}{\bibfnamefont{N.}~\bibnamefont{Vanhaecke}},
  \bibinfo{author}{\bibfnamefont{H.~L.} \bibnamefont{Bethlem}},
  \bibnamefont{and} \bibinfo{author}{\bibfnamefont{G.}~\bibnamefont{Meijer}},
  \bibinfo{journal}{Phys. Rev. A}
  \textbf{\bibinfo{volume}{71}}(\bibinfo{number}{5}), \bibinfo{eid}{053409}
  (\bibinfo{year}{2005}{\natexlab{c}}).

\bibitem[{\citenamefont{K\"upper et~al.}(2006)\citenamefont{K\"upper, Haak,
  Wohlfart, and Meijer}}]{Kuepper:RSI77:016106}
\bibinfo{author}{\bibfnamefont{J.}~\bibnamefont{K\"upper}},
  \bibinfo{author}{\bibfnamefont{H.}~\bibnamefont{Haak}},
  \bibinfo{author}{\bibfnamefont{K.}~\bibnamefont{Wohlfart}}, \bibnamefont{and}
  \bibinfo{author}{\bibfnamefont{G.}~\bibnamefont{Meijer}},
  \bibinfo{journal}{Rev. Sci. Instrum.}
  \textbf{\bibinfo{volume}{77}}(\bibinfo{number}{1}), \bibinfo{pages}{016106}
  (\bibinfo{year}{2006}).

\bibitem[{\citenamefont{van~de Meerakker
  et~al.}(2006{\natexlab{b}})\citenamefont{van~de Meerakker, Vanhaecke, and
  Meijer}}]{Meerakker:ARPC57:159}
\bibinfo{author}{\bibfnamefont{S.~Y.~T.} \bibnamefont{van~de Meerakker}},
  \bibinfo{author}{\bibfnamefont{N.}~\bibnamefont{Vanhaecke}},
  \bibnamefont{and} \bibinfo{author}{\bibfnamefont{G.}~\bibnamefont{Meijer}},
  \bibinfo{journal}{Ann. Rev. Phys. Chem.} \textbf{\bibinfo{volume}{57}},
  \bibinfo{pages}{159} (\bibinfo{year}{2006}{\natexlab{b}}).

\bibitem[{\citenamefont{Fulton et~al.}(2006)\citenamefont{Fulton, Bishop,
  Shneider, and Barker}}]{Fulton:NatPhys2:465}
\bibinfo{author}{\bibfnamefont{R.}~\bibnamefont{Fulton}},
  \bibinfo{author}{\bibfnamefont{A.~I.} \bibnamefont{Bishop}},
  \bibinfo{author}{\bibfnamefont{M.~N.} \bibnamefont{Shneider}},
  \bibnamefont{and} \bibinfo{author}{\bibfnamefont{P.~F.}
  \bibnamefont{Barker}}, \bibinfo{journal}{Nature Physics}
  \textbf{\bibinfo{volume}{2}}, \bibinfo{pages}{465} (\bibinfo{year}{2006}).

\bibitem[{\citenamefont{Meek et~al.}(2008)\citenamefont{Meek, Bethlem, Conrad,
  and Meijer}}]{Meek:PRL100:153003}
\bibinfo{author}{\bibfnamefont{S.~A.} \bibnamefont{Meek}},
  \bibinfo{author}{\bibfnamefont{H.~L.} \bibnamefont{Bethlem}},
  \bibinfo{author}{\bibfnamefont{H.}~\bibnamefont{Conrad}}, \bibnamefont{and}
  \bibinfo{author}{\bibfnamefont{G.}~\bibnamefont{Meijer}},
  \bibinfo{journal}{Phys. Rev. Lett.} \textbf{\bibinfo{volume}{100}},
  \bibinfo{pages}{153003} (\bibinfo{year}{2008}).

\bibitem[{\citenamefont{Courant and Snyder}(1952)}]{Courant:AnnPhys3:1}
\bibinfo{author}{\bibfnamefont{E.~D.} \bibnamefont{Courant}} \bibnamefont{and}
  \bibinfo{author}{\bibfnamefont{H.~S.} \bibnamefont{Snyder}},
  \bibinfo{journal}{Phys. Rev.} \textbf{\bibinfo{volume}{88}},
  \bibinfo{pages}{1190} (\bibinfo{year}{1952}).

\end{thebibliography}
\end{document}